\documentstyle[12pt,aasms4]{article}

\received{}
\accepted{}
\journalid{}{}
\articleid{}{}

\slugcomment{ to appear in the Astrophysical Journal}

\lefthead{Krennrich et al.}
\righthead{TeV Flare-Spectra of Markarian 421 and 501}

\begin{document}

\def\lsim{\mathrel{\hbox{\rlap{\hbox{\lower4pt\hbox{$\sim$}}}\hbox{$<$}}}}
\def\gsim{\mathrel{\hbox{\rlap{\hbox{\lower4pt\hbox{$\sim$}}}\hbox{$>$}}}}

\title{Measurement of the Multi-TeV $\gamma$-ray Flare-Spectra 
of \\ Markarian 421 and Markarian 501  }

\author{F.~Krennrich\altaffilmark{1}, S.~D.~Biller\altaffilmark{2}, I.~H.~Bond\altaffilmark{3}, 
P.~J.~Boyle\altaffilmark{4}, S.~M.~Bradbury\altaffilmark{3}, A.~C.~Breslin\altaffilmark{4},
J.~H.~Buckley\altaffilmark{5}, A.~M.~Burdett\altaffilmark{3}, J.~Bussons
Gordo\altaffilmark{4}, D.~A.~Carter-Lewis\altaffilmark{1},
M.~Catanese\altaffilmark{1}, M.~F.~Cawley\altaffilmark{6},
D.~J.~Fegan\altaffilmark{4}, J.~P.~Finley\altaffilmark{7},
J.~A.~Gaidos\altaffilmark{7}, T.~Hall\altaffilmark{7},
A.~M.~Hillas\altaffilmark{3}, R.~C.~Lamb\altaffilmark{8}, 
R.~W.~Lessard\altaffilmark{7}, C.~Masterson\altaffilmark{4}, 
J.~E.~McEnery\altaffilmark{9}, G.~Mohanty\altaffilmark{1,10},
P.~Moriarty\altaffilmark{11}, J.~Quinn\altaffilmark{12},
A.~J.~Rodgers\altaffilmark{3}, H.~J.~Rose\altaffilmark{3}, 
F.~W. Samuelson\altaffilmark{1},  G.~H.~Sembroski\altaffilmark{6},
R.~Srinivasan\altaffilmark{6}, V.~V.~Vassiliev\altaffilmark{12},
T.~C.~Weekes\altaffilmark{12}}

\altaffiltext{1}{Department of Physics and Astronomy, Iowa State
University, Ames, IA 50011-3160}

\altaffiltext{2}{Department of Physics, Oxford University, Oxford, UK}

\altaffiltext{3}{Department of Physics, University of Leeds,
Leeds, LS2 9JT, UK}

\altaffiltext{4}{Experimental Physics Department, University College, 
Belfield, Dublin 4, Ireland}

\altaffiltext{5}{Department of Physics, Washington University, St.~Louis,
MO 63130}

\altaffiltext{6}{Physics Department, National University of
Ireland, Maynooth, Ireland}

\altaffiltext{7}{Department of Physics, Purdue University, West
Lafayette, IN 47907}

\altaffiltext{8}{Space Radiation Laboratory, California Institute of
Technology, Pasadena, CA 91125}

\altaffiltext{9}{Present address: Department of Physics, University of Utah,
Salt Lake City, UT 84112}

\altaffiltext{10}{Present address: LPNHE  Ecole Polytechnique, 91128 Palaiseau CEDEX, France } 

\altaffiltext{11}{School of Science, Galway-Mayo Institute of Technology, Galway, Ireland}

\altaffiltext{12}{ Fred Lawrence Whipple Observatory, Harvard-Smithsonian 
CfA, P.O.~Box 97, Amado, AZ 85645-0097}

\clearpage
\begin{abstract}

The energy spectrum of Markarian~421 in flaring states has been
measured from 0.3 to 10 TeV using both small zenith angle and large
zenith angle observations with the Whipple Observatory 10 m imaging
telescope.  The large zenith angle technique is useful for extending
spectra to high energies, and the extraction of spectra with this
technique is discussed.  The resulting spectrum of Markarian~421 is
fit reasonably well by a simple power-law: $ \rm J(E) = \: \:
E^{-2.54 \: \pm \: 0.03 \: \pm \: 0.10} \: photons \: m^{-2} \: s^{-1}
\: TeV^{-1} $ where the first set of errors is statistical and the
second systematic.  This is in contrast to our recently reported spectrum of
Markarian 501 which, over a similar energy range, has substantial
curvature.  The differences in TeV energy spectra of gamma-ray blazars
reflect both the physics of the gamma-ray production mechanism and possibly
differential absorption effects at the source or in the intergalactic
medium.  Since Markarian~421 and Markarian~501 have almost the same redshift
(0.031 and 0.033 respectively), the difference in their energy spectra must be
intrinsic to the sources and not due to intergalactic absorption,
assuming the intergalactic infrared background is uniform.

\end{abstract}


\section{Introduction}

Three blazars Markarian~421 (Punch et al.\ 1992), Markarian~501
(Quinn et al.\ 1996) and 1ES 2344+514 (Catanese et al.\ 1998) have
been detected at TeV energies.  At GeV energies, the EGRET instrument
aboard the Compton Gamma-Ray Observatory found only upper limits for
Markarian~501 (Catanese et al. 1997) and 1ES 2344+514 (Thompson 1996).
Markarian~421 was detected by EGRET, but only very weakly 
(Thompson et al. 1995).  While blazar emission for X-ray selected objects
at lower energies (up to about 1-100 keV) is almost certainly
due to synchrotron emission from a beam of highly relativistic
electrons, the GeV/TeV emission forms a second component usually
attributed to inverse Compton scattering of relatively low energy
photons by the electron beam (see, e.g., Sikora, Begelman \&
Rees 1994) or perhaps to pion photoproduction by a proton component of
the beam (see, e.g., Mannheim \ 1993).  The inverse Compton
models predict typical blazar gamma-ray energy cutoffs of 10 GeV to
about 30 TeV whereas proton beam models allow gamma-ray
energies exceeding 100 TeV.  Spectrum measurements at the low energy
end (perhaps probing the energy threshold of the second component) and
the high energy end (perhaps showing an energy cutoff) are both important
for constraining models.

The TeV gamma-ray spectra of extra-galactic objects can be
modified by differential absorption due to photon-photon collisions
with inter-galactic IR radiation (Gould \& Schr\` eder 1967; Stecker, de
Jager \& Salamon 1992).  Indeed, TeV observations of Markarian~421 and
Markarian~501 (Zweerink et al.\ 1997; Krennrich et al. 1997; Aharonian
et al.\ 1997) have been used to set upper limits on the density of intergalactic
IR radiation (Stanev \& Franceschini 1998; Biller et al.\ 1998).  
These upper limits provide the best constraints on infrared densities 
in the 0.02 - 0.2 eV regime and do not suffer from local galactic background 
contributions as are present in direct measurements. At this time, no unambiguous
evidence has been found for an IR absorption spectral cutoff.  In
order to infer the magnitude of intergalactic IR background radiation
from absorption effects on TeV spectra, it is necessary to have
a good model for intrinsic spectra or to have spectra from 
several objects and assume that the intrinsic TeV spectra
of the objects are identical (or at least very similar), or to have
detections from many sources and assume that they are similar in a
statistical sense.  For the redshift range (z = 0.031-0.044) of the
detected TeV blazars Markarian~421, Markarian~501 and 1ES~2344+514, a
recognizable cutoff (optical depth $\sim$ 1-5) is expected to occur
between 5-20 TeV (Stecker et al.\ 1997; Stecker et al. 1998).

We have recently published a brief report giving a detailed spectrum
of Markarian~501 spanning the energy range 260 GeV to 10 TeV derived
from observations with the Whipple Observatory gamma-ray imaging
Cherenkov telescope (Samuelson et al.\ 1998).  This spectrum was
derived from observations made during ``high'' states of the AGN which
give good statistical precision.  Markarian~501 (Quinn et al.\ 1998)
is variable in the TeV energy range showing changes on a time scale of
several hours.   The data were taken at both standard small zenith angles
(SZA) of less than 25 degrees and at large zenith angles (LZA) in the
range of 55 to 60 degrees.  The SZA observations are sensitive to
relatively low energies ($\rm E < $ 5 TeV) and the LZA observations yield 
better statistics at high energies.  
The details of the analysis were not explained in the brief
report but are given here.  In particular, the characteristics of
Cherenkov imaging telescopes relevant for spectral determination
change substantially from SZA to LZA observations.  In addition,
uncertainties in the spectrum, e.g., due to corresponding
uncertainties in atmospheric absorption, are also given here.
Finally, we have used the flux from the Crab Nebula (which often serves as a 
standard candle in TeV gamma-ray astronomy) to check spectra extracted from
LZA observations against our standard spectrum (Hillas et al.~1998a).

These recent results from Markarian~501 indicate a spectrum which is not 
consistent with a simple power-law in which the flux, J, is proportional 
to E$^{-\gamma}$, but is more accurately described by a three parameter 
curve (parabolic in a plot of $\log(\hbox{J})$ vs.\ $\log(\hbox{E})$) 
given by 
$\hbox{J} \sim \hbox{E}^{-2.22\pm 0.04 \pm 0.05-(0.47\pm 0.07)\log_{10}(\hbox{E})}$
where the first set of errors is statistical and the second systematic
and E is in TeV.  (Previously published spectra of Markarian~501
(Bradbury et al.\ 1997; Aharonian et al. 1997) covered a smaller
energy span and were consistent with a simple power-law.)  In
principle, the curvature of the spectrum could arise either from the
physics of gamma-ray emission from AGN, from intrinsic absorption or 
from absorption in the inter-galactic medium.

In addition to Markarian~501, the Whipple Observatory Gamma-ray
collaboration has a database of both SZA and LZA data for Markarian
421.  Fortuitously, these objects have almost identical redshifts:
0.031 for Markarian 421 and 0.033 for Markarian 501.  Hence any 
differences in their TeV
spectra must be intrinsic to the AGN and not due to intergalactic
absorption assuming the intergalactic IR background radiation is
uniform.  Like Markarian 501, Markarian 421 is also highly
variable, with changes observed on a time scale as short as 15 minutes
(Gaidos et al.\ 1996).  We have previously published a TeV spectrum
for Markarian~421 based on a single SZA observation, but very high
flux detection lasting only 2 hours (Zweerink et al.~1997).  The
published spectrum was consistent with a simple power-law i.e., no
curvature was required for an acceptable fit.  However the spectrum
did not cover as large an energy range as that for Markarian~501, nor
did the Markarian~421 spectrum have comparable statistical precision
at the high energy end of the spectrum.   Hence it was not possible 
to draw firm conclusions from a comparison of the two spectra.

Here we present a new TeV spectrum of Markarian~421 based upon both SZA
and LZA data taken while the AGN was in a high state of emission.  The
SZA data set used consists of two flaring detections.  The first is the
same 2-hour detection (average flux of 7.4 Crab units) used to obtain
the previously published spectrum (Zweerink et al.~1997), but we have 
taken additional care in the treatment of energy threshold effects in order 
to obtain flux values at lower energies.  The second detection (27 minutes on-source)
had an average flux of 2.8 Crab units and exhibited a remarkably short
rise and fall time of 15 minutes.  It occurred 8 days after the first
detection. The LZA observations consisted of 1.5 hours (1995 June) of on-source 
data at zenith angles of 55 to 60 degrees with an average flux of 3.3 Crab units.
The new spectrum is consistent with our previously published result,
but spans a larger energy range (comparable to that published for
Markarian~501) and has better statistics at high energies.  The
spectrum appears less curved than the Markarian~501 spectrum.
We show that the spectra for the two equidistant AGN (Markarian~421
and Markarian~501) clearly differ, and this reflects intrinsic spectral
differences near the sites of gamma-ray production.  Combined with
data for the two AGN at X-ray energies, these spectra constrain models
of physical processes in the jets (see Hillas et al.~1998b).  The
differences also point toward a major difficulty in inferring
intergalactic background radiation intensities via TeV photon
attenuation.

\section{Observations}

The observations presented here were made with the Whipple 
Observatory 10 m imaging Cherenkov telescope.  
The camera consisted of 109 (until 1996 December 4) or 151 (after 1996 December 4)
photomultiplier tubes (PMTs) placed on a 1/4 degree hexagonal matrix.  These
cameras covered fields of view of about 2.7 and 3.2 degrees
respectively.  Pulses from each of the photomultiplier tubes were
amplified and sent to gated integrating adc's (analog-to-digital converters)
 and, in addition, those from the inner 91 tubes 
were sent to discriminators with levels
corresponding to about 49 photoelectrons.  When two of the
discriminators fired, a trigger was generated, and current pulses from
all photomultiplier tubes were integrated for 25 nanoseconds and
digitized (Cawley et al.~1990).  The data were normally taken in an
on-off mode in which the source is tracked for typically 28 minutes, and
then the same range of elevation and azimuth angles is tracked for
another 28 minutes giving a background comparison region.  The Crab
Nebula serves as a standard candle for TeV gamma-ray astronomy, and it
was observed using the same camera configurations and ranges of zenith
angles that were used for the blazar observations.

As described previously in Samuelson et al.~(1998), the Markarian 501
observations were made in the time interval from February 14 to June 8
of 1997 during a high state of the source in 1997 (Protheroe et
al. 1997).  During this period the camera had 151 pixels giving the
larger (3.2 degree) field of view.  A total of 16 hours of SZA
on-source data were taken at 8 to 25 degrees and 5.1 hours of LZA
on-source data were taken at 55 to 60 degrees.  These observations
showed that the flux for the 1997 observing season varied from about 0.2
to 4 times the flux from the Crab Nebula with an average value of 1.4
(c.f. Quinn et al. 1998).  This is a factor of 7 larger than the
average flux from the 1996 observing season, which is the basis for
identifying it as flare data.  We are somewhat arbitrarily defining
flare data as that which has a flux level substantially above the
average flux as measured for a given source in our observations.

The SZA data for Markarian 421 consists of two detections on 1996 May 7 and
May 15, measured with the 109-photomultiplier tube camera.  The
first of these was during the highest flux TeV flare observed.  It
consisted of a 2 hours observation in which the data rate increased
steadily giving a count rate at the end of the run of about 15~gammas/minute 
at $\rm E > $ 350 GeV which is 10 times the rate from the Crab Nebula.  The
average rate during the runs was 7.4~Crab units.  During the flare the
AGN was observed continuously, and hence background comparison regions
were taken from the same range of elevation and azimuth angles but
from other nights.  However, because of the
strength of the signal, the data are almost free of background after
selection of gamma-ray-like images.  The results are insensitive to
the exact background runs used (Zweerink et al.~1997; Zweerink
1997).  The second detection eight days later consisted of 27 minutes
of on-source observations with corresponding off-source data.  The
average flux was 2.8 Crab units and the data show a remarkable peak
with a rise and fall time of only 15 minutes.  The Crab Nebula
database was measured with the same camera during the 1995/96 season
and consists of 49 on/off pairs giving the Crab Nebula spectrum
reported by Hillas et al.~(1998a).

Observations of Markarian~421 at LZA were carried out in 1995 with the
109 pixel camera (Krennrich et al.~1995), and the detection of 5-8 TeV gamma rays was
reported earlier (Krennrich et al.~1997).  For the spectral analysis
presented here, we used a subset of the data for which the range of zenith angles 
is 55-60 degrees where we have adequate observations of the Crab
Nebula to obtain a spectrum.  This allows us to use the Crab Nebula to
test the LZA analysis procedure and show that it is consistent with
spectra derived at SZA.  In addition, we required that Markarian 421 was in a
flare state, which gave a total of three on-off pairs of data measured
1995 on June 20, 29 and 30 with an average flux from the AGN of 3.3
Crab units, comparable to that of the May 15 flare (an average of 2.8
Crab units).

Data from observations of the Crab Nebula at zenith angles between 55
and 60 degrees were collected during 1995/1996/1997 using both the
109-pixel and 151-pixel cameras.  A total of 17~on-off pairs (8 hours
on-source) were used for the derivation of the energy spectrum of 
the Crab Nebula.  The spectra derived from the two cameras were in agreement.  

\section{Extraction of Spectra from SZA and LZA Data}

For gamma-rays with energies of a few hundred GeV incident near the
zenith, shower maximum ($shower$~$max$: the region along the 
longitudinal development of the electromagnetic cascade with the 
maximum number of electrons and positrons) occurs at about 7-9 km 
above sea level.  The Cherenkov light from such a shower forms a 
pool of light about 200~m in diameter
at telescope altitude, 2.3~km for the Whipple Observatory.  At large
zenith angles, $shower$~$max$ occurs farther away from the telescope, 
increasing the area
over which the Cherenkov light is distributed.  The lower light
density raises the telescope energy threshold, but for gamma rays with
sufficient energy to produce enough light for triggering, the
collection area is substantially larger (Sommers \& Elbert 1987).
Since $shower$~$max$ occurs at a high altitude for LZA showers, the 
characteristic Cherenkov angles are smaller resulting in a smaller 
Cherenkov light image nearer the center of the field of view of the 
camera.

In this work we followed an established Whipple procedure in extracting
spectra, specifically Method I as described in detail in Mohanty et
al.~(1998).  The following parts are required: (1) a method for
selecting gamma-ray initiated shower images from a background of
cosmic-ray initiated shower images based upon image shape and
orientation, (2) the effective telescope collection area for this selection
method, (3) a method to estimate the initial gamma-ray energy for each
event, and (4) the resolution function corresponding to this energy
estimate.  The method for selecting gamma-ray events should be
relatively independent of the gamma-ray energy, E.  The method of energy
estimation should give good resolution and be relatively free of bias.
These parts are described in the next two sections and, in the last
subsection, we show that spectra derived from Crab Nebula LZA
observations agree with our standard SZA results published earlier.

\subsection{Energy Threshold, Selection Criteria and Collection Area}

The Cherenkov imaging technique utilizes differences in focal plane
image shapes to differentiate cosmic-ray background from gamma rays.
The selection criteria (Mohanty et al. 1998) are based on image 
shape through the $width$
and $length$ parameters, and on orientation through the $alpha$
parameter.  Compared with cosmic-ray images gamma-ray images 
are generally narrower, shorter and point toward the center 
of the focal plane (see, e.g., Hillas~1985; Reynolds et al.~1993).
The effective collection area of an imaging atmospheric Cherenkov
telescope is limited not only by the dimension of the Cherenkov light
pool on the ground but also by the image parameter cuts which are applied
to increase the signal to noise ratio.  The criteria are derived from the
parameter distributions of simulated gamma-ray showers as a
function of their total light intensity ($size$, hereafter) 
in the photomultiplier camera.  We set these criteria so that they keep
90 \% of gamma-ray images whose centroid is 0.5-1.1 degrees ($distance$) 
from the center of the field of view for the 151-pixel
camera.  This distance restriction improves the correlation of energy with
$size$.  The cuts are scaled with $size$ so that the efficiency for
keeping gamma rays is approximately independent of energy. 

The telescope is triggered when two of the inner 91 photomultiplier
tubes give pulses within a triggering gate of 30 ns
with 49 or more
photoelectrons.  The trigger electronics (Cawley et al.~1990) is
difficult to model precisely.  One of the problems is that the
photomultiplier tube pulses include both Cherenkov light ``signal''
and Poisson-fluctuating night-sky noise ``background'' which causes
the pulse shapes to vary.  The pulses go both to a discriminator which
fires when the pulse voltage crosses a preset threshold and to an
integrating analog-to-digital converter which records the total
charge, $q$, in the pulse, (c.f. Mohanty~1995).  Because of the variation
in pulse shape, there is no unique correspondence between pulse charge
and peak voltage effectively giving the discriminators a ``fuzzy''
edge having a width corresponding to 3.5 photoelectrons about a mean
trigger point of 49 photoelectrons.  In addition, if the discriminator
levels are set very low, the background trigger rate for low-light
events can be sensitive to night-sky brightness.

We avoid these difficulties using software padding (Cawley 1993) 
and by adding the additional software requirement that a
signal corresponding to at least $\rm D_{soft} = 80$ photoelectrons
is present in at least two pixels.  This raises the
telescope energy threshold, but the collection area can be readily
calculated. 

The resulting SZA (20 degrees) and LZA (55-60
degrees) telescope areas for events which pass both the triggering and
image-selection requirements are shown in Fig.~1 for the 151-pixel
camera.  It is clear that only SZA measurements have sensitivity below
1~TeV whereas the LZA measurements have better sensitivity
beyond about 5~TeV.  The LZA collection area shows a plateau between
about 3~TeV and 50~TeV.  There is a SZA/LZA overlap region for cross
calibration between about 1 and 10~TeV.

One concern is that we have properly extracted SZA spectra at all but the
lowest energy point in Fig.~1.  The points at 260 GeV and 380 GeV have
significantly reduced collection area and hence might be unusually
sensitive to small details in the simulations.  We have looked for
such sensitivities by varying the (1) the telescope gain, (2)
reflector optical resolution and (3) background sky noised used in the
simulations.  The result is that neither the calculated collection
area nor the extracted spectra change significantly if these
parameters are varied over physically reasonable values.  (Indeed this
is a basis for arriving at systematic errors.)  Furthermore, the
gamma-ray image parameter distributions extracted using on-off
histograms (see Mohanty et al., 1998) agree with simulations.  Thus
the results appear to be robust.

\subsection{Energy Estimation and Resolution}

The accuracy of the energy reconstruction of gamma-ray primaries with
a single imaging Cherenkov telescope is limited by the following
effects: a) fluctuations in the first interactions which cause the height of
shower maximum to vary (hence the region where most of
the Cherenkov light is emitted varies causing fluctuations in the
light density detected at ground), b) the uncertainty in the shower core
distance to the telescope, and  c) truncation of the shower images close to
the edge of the field of view.

All three effects occur in a similar fashion for SZA and LZA
observations. However, there are some differences: the central light
emitting region for a LZA shower appears geometrically smaller in the
camera, because of its larger distance from the instrument and the
smaller Cherenkov angles in the lower density atmosphere.  Hence 
a bigger fraction of the Cherenkov light image is contained in the 
field of view.  Also, the smaller Cherenkov angles for LZA observations shift the
center of gravity of images closer to the center of the field of
view.  Therefore, truncation effects are less important for LZA data.

Following Mohanty et al.~(1998), we have found expressions for an energy
estimate, $\rm \widetilde E$ as a function of $\ln(size)$ and
$distance$ which are relatively free of bias and have good resolution.  To a good
approximation, for fixed energy, E, the distribution $\rm \widetilde
E$ is lognormal (see Mohanty et al.~1998) with a width independent of E.
It follows that the telescope resolution $\Delta\hbox{E}$ in standard
deviations is given by $\Delta\hbox{E}/\hbox{E} = \sigma$ where
$\sigma$ is about 0.34 for SZA and slightly better, about 0.29 for
LZA.

\subsection{Atmospheric Effects}

Since Cherenkov light from showers at LZA passes through substantially
more atmosphere than at SZA, the uncertainties in the atmospheric
model may have correspondingly larger effects.  There are four
important extinction mechanisms.  (1) The largest is Rayleigh
scattering for which the cross-sections are well known.  Variations
arise because of changes in barometric pressure (typically a few
percent) changing the column density along the line of sight from the
telescope to shower maximum.  (2) Ozone exists mainly at altitudes
well above shower maximum, but the
cross sections for UV absorption are very large and small
concentrations extend into lower regions causing some absorption.
Seasonal variations are of order 20-25\% at the Whipple Observatory
latitude and there are daily variations as well.  (3) Absorption by
$\hbox{O}_2$ becomes important below about 250 nm and removes almost
all the light at shorter wavelengths.  There are significant
uncertainties in the absorption cross sections, but these
uncertainties do not have a large effect because the absorption turns
on rapidly and essentially all the light below 250 nm is absorbed in
any case.  (4) Aerosols exist mainly at low altitudes with a scale
height of roughly 1 km, and hence the observatory altitude of 2.3 km
diminishes their effects.  One of their primary characteristics is
variability.  If aerosol absorption were significant, one would expect
significant variability in the telescope cosmic-ray induced trigger
rate, whereas this is usually stable to a few percent.

In order to estimate the effects of atmospheric uncertainties, we made
some simple calculations using the atmospheric model used for the ARTEMIS
project (Urban et al.~1996) and a simple aerosol parameterization used
for the Fly's Eye experiment (Baltrusaitis et al.~1985).  Assuming a
Cherenkov light spectrum with wavelength-dependent mirror reflectivity
and photocathode quantum efficiency folded in, the transmission for
light from $shower$~$max$ (for 5 TeV $\gamma$-rays) to the telescope was calculated
under various assumptions.  The altitude of shower maximum for showers
initiated by gamma rays at the zenith is 9 km and 10 km for gamma rays
incident at zenith angles of 60 degrees.  The results are given in
Table~1: row 1 corresponds to standard atmospheric conditions; row 2
has the Rayleigh scattering column density increased by 3\% 
(due to barometric pressure changes); row 3 has
the aerosol concentration increased by a factor of 4; row 5 has the
$O_2$ cross sections increased by a factor of 4; and row 6 has the
ozone concentration increased by a factor of 4.  As can be seen from
the table, atmospheric transmission from shower maximum at 60 degrees is
78 \% of that at the zenith. However, changes in transmission due to
fairly large increases in various extinction mechanisms is on the
order of few percent. This is small compared with the overall
uncertainties in telescope gain of about 15 \% (Mohanty et al. 1998).  

\subsection{The Crab Nebula Spectrum from LZA Data}

As a check on extraction of spectra from LZA observations, we have
analyzed existing data for the Crab Nebula (1995-1997 with the 109
and the 151 pixel camera).  In the angular range of 55-60 degrees, this consists
of 8 hours of on-source data with corresponding off-source runs.  When
analyzed as described above, the resulting spectrum is shown in
Fig.~2.  Also shown in the figure is the standard spectrum as given in
Hillas et al.~(1998a) derived there using SZA observations.  It is
apparent that the two agree over the common energy range of about 1.1
to 10 TeV.  A power-law fit to LZA data yields: $ \rm J(E) = 3.78 \:
\times \: 10^{-7} \: E^{-2.59 \: \pm \: 0.15
\:  \pm \:  0.15}  \: photons \:  m^{-2}  \: s^{-1}   \: TeV^{-1} $.
The first set of uncertainties are statistical and the second are systematic 
calculated as in Mohanty et al. (1998). Based on the Crab analysis
and on our simulations of the effect of variations in the atmospheric
model we are confident about the SZA and the LZA energy estimate.

\section{The Markarian 421 Spectrum}

We have reanalyzed the SZA flare data of 1996 May 7, using a
more careful treatment of the telescope threshold region.  The new
spectrum is consistent with our previously published result, but now
includes two lower energy points extending down to 260 GeV instead of 560 GeV.
The SZA flare data of 1996 May 15, were analyzed in exactly the same
way.  As pointed out previously, the threshold region is difficult to
model, and to avoid it in the previous analysis, we imposed a
secondary software trigger level by requiring that two of the
triggering tubes have signals corresponding to at least 50
photoelectrons ($\rm D_{soft}$ = 49) and that the $size$ was at least
400 photoelectrons.

In the present analysis, we impose no direct limitation on the signal
$size$, but instead require that $\rm D_{soft} = 80$.  This also
avoids the troublesome region, but with less cost in energy threshold.
We have studied the effect on varying $\rm D_{soft}$ and found that
the flux values in the spectrum are stable above a value of about 70.
As $\rm D_{soft}$ is increased above 100, the flux values do not
change within statistical errors, but these errors become
significantly larger.  We have also investigated another trigger
configuration in which at least three tubes were required to have $\rm
D_{soft}$ greater than 80.  This again led
to the same flux values within errors. 

Finally, we reanalyzed the 1994/1995 Crab database and found that it is 
well fit over the energy range 0.3 to 10 TeV with a simple power-law 
consistent with the previous result given by Hillas et al.~(1998a).
 
The spectral flux values derived from the intense flare of Markarian~421
at SZA are shown as stars in Fig.~3.  The data are fit by: 

$ \rm J(E) = 2.2 \:
\times \: 10^{-6} \: E^{-2.54 \: \pm \: 0.04
\:  \pm \:  0.1}  \: photons \:  \: m^{-2}  \: s^{-1}   \: TeV^{-1} $
giving a $\chi^2$ of 22.8 for 8 degrees of freedom (probability
0.005).  This fit is marginal, perhaps indicating some curvature. The
May 15, 1996 data are shown in the same figure as boxes and are fit
by: $ \rm J(E) = 1.0 \:
\times \: 10^{-6} \: E^{-2.45 \: \pm \: 0.10
\:  \pm \:  0.1}  \: photons \:  m^{-2}  \: s^{-1}   \: TeV^{-1} $
giving a $\chi^2$ of 3.2 for 7 degrees of freedom.  The LZA energy
spectrum covers 1.5-10.4 TeV and is shown as open circles in the same
figure. The LZA points can be fitted by a power-law of the form: $ \rm
J(E) = 7.53 \: \times \: 10^{-7} \: E^{-2.52 \: \pm \: 0.18
\:  \pm \:  0.15}  \: photons \: \: m^{-2}  \: s^{-1}  \: TeV^{-1} $
giving a $\chi^2$ of 4.9 for 4 degrees of freedom.

Since all the spectral shapes are consistent we combine them in order to 
reduce the statistical uncertainties.
In combining the two SZA data sets and the LZA data shown in Fig.~4,
the normalizations of the May 15 SZA data and the LZA were treated as
free parameters thus fixing the absolute normalization to the May 7
flare.  The resulting fit is: $ \rm J(E) \propto \: E^{-2.54 \:
\pm \: 0.03 \: \pm \: 0.10} \: photons \: m^{-2} \: s^{-1} \: TeV^{-1} $
giving a $\chi^2$ of 31.5 for 21 degrees of freedom and a chance
probability of 0.07.  Thus, the energy spectrum of Markarian~421
between 260 GeV - 10 TeV during flaring activity is
consistent with a single power-law.  A curved fit for Markarian 421 yields: 
$$ \hbox{J(E)} =(2.4\pm 0.1\pm0.3)\times
10^{-6} (\frac{\hbox{E}}{1
\hbox{TeV}})^ {-2.47\pm 0.04 \pm 0.05-(0.28\pm
0.09)\log_{10}(\rm E)}$$  $\rm photons \:  m^{-2}  \: s^{-1}  \: TeV^{-1} $ 
with a $\chi^2$ value of 21.5 for 20 degrees of freedom giving a chance 
probability of 0.4.

\section{The Markarian 501 Spectrum} 

The Markarian 501 spectrum was analyzed in a similar way.  The
results for the 5.1 hours of LZA observations are shown together with
the LZA spectrum for the Crab Nebula in Fig.~5.  The flux level of
Markarian~501 was on average $\approx$ 2 Crab units during these
observations and the spectral slope is
similar to that for the Crab spectrum.  The spectrum extends up to 10
TeV and can be fit between 1.1-10.4 TeV with a power-law yielding
$\chi^{2} = 14.7$ for 5 degrees of freedom (chance probability of
0.015):
 
$ \rm J(E) = 7.53 \: \times \: 10^{-7} \: E^{-2.67 \: \pm \: 0.09 \:
\pm \: 0.15} \: photons \: m^{-2} \: s^{-1} \: TeV^{-1} $.
 The errors on the spectral index are given by a statistical
uncertainty of $\pm \: 0.09$, and a systematic uncertainty of $\pm \:
0.15$.  The slightly high value of $\chi^2$ hints at curvature in the
spectrum.  

The LZA data (5.1 hours) can be combined with the SZA data (15 hours)
treating the normalization of the former as a free parameter as
described in the last section.  This yields the spectrum given
previously in Samuelson et al.~(1998) which is shown in Fig.~6.  Fitting
this data with a simple power-law:

$ \rm J(E) = 6.9 \:  \times  \: 10^{-7}  \:  E^{-2.41 \:  \pm \:  0.025
\: }  \: photons \:  m^{-2}  \: s^{-1}   \: TeV^{-1} $,

\noindent giving $\chi^2 = 59.7$ for 15 degrees of freedom with a chance
probability of $2.5 \times 10^{-7}$.  Including a curvature term
yields

$ \rm J(E) = (8.6 \pm 0.3 \pm 0.7) \: \times \: 10^{-7} \: E^{-2.22
\: \pm \: 0.04 \pm \: 0.05 \: -(0.47 \pm 0.07)log_{10}(E) } \: photons
\: m^{-2} \: s^{-1} \: TeV^{-1} $.

\noindent giving $\chi^2 = 18$ for 14 degrees of freedom with a chance
probability of 0.2.  As shown in Fig.~6, the Markarian 501 spectrum
is clearly curved.  Spectral variability is not likely to account for
the curvature, the superposition of two different power-laws would result
in a concave spectrum rather than a convex shape.

\section{Discussion: Markarian 421 vs. Markarian~501}

The spectra derived from LZA and SZA data for Markarian 421 and Markarian 501
are shown in Fig.~7.  It is apparent from the figure that they
differ; a $\chi^2$ test places the chance probability that they arise
from the same parent distribution at $ \rm 4 \times 10^{-3} $.  We
conclude that the energy spectra of Markarian 421 and Markarian 501
during flaring activity are different.

Although Markarian 421 and Markarian 501 are at almost the same redshift, they
do differ in their X-ray spectrum. Observations of Markarian 421 by
the ASCA X-ray satellite experiment, although not contemporaneous with
the data presented here, indicate an energy break in the synchrotron 
spectrum of 1.6~-~2.2~keV (Takahashi et al. 1996).  In contrast, 
X-ray observations of
Markarian 501 by BeppoSAX taken in April 1997 showed that its
synchrotron power can peak at hard X-ray energies at 100~keV (Pian et
al.~1997). These observations coincide with long term flaring activity
in TeV gamma rays (February to August 1997) and indicated that synchrotron
 power from an AGN can peak at hard X-ray energies, beyond 100~keV.  
In addition Markarian~501 has been detected by the OSSE
instrument aboard the Compton Gamma-ray Observatory at energies of
50-470~keV (Catanese et al. 1997) showing clearly, for the first
time, that synchrotron emission can peak above 100 keV.

At GeV energies, Markarian~421 is seen by EGRET (Lin et al. 1992),
albeit weakly, whereas Markarian~501 is not (Catanese et al. 1997).  Thus, in
terms of the synchrotron-inverse Compton models for which the GeV
emission is from the inverse-Compton mechanism, it would appear that
both the synchrotron peak and the inverse-Compton peak are shifted to
higher energies leaving the EGRET GeV energy sensitivity range in the
gap between them for Markarian~501.  As shown in Fig.~7, in the energy
range 260 GeV~-~10~TeV, the spectrum of Markarian~501 is harder at lower
energies and shows more curvature than Markarian~421. In fact the latter is 
consistent with a straight line (i.e., pure power-law).  This is
also consistent with the peak inverse-Compton power occuring at higher
energies for Markarian~501, nearer the range covering our
measurements.  We see no obvious contradiction of our results with a
synchrotron-inverse Compton picture for the origin of the TeV
radiation.

In order to probe inter-galactic IR radiation via attenuation of TeV
gamma rays, it is first necessary to know the intrinsic energy spectra
of AGN.  Spectral features such as the curvature of Markarian 501
cannot be ascribed a priori to this attenuation mechanism.  This is
clear because Markarian 421 and Markarian 501 have almost
identical redshifts yet different spectra. The differences in
their spectra can perhaps be explained in the context of the
synchrotron-inverse Compton picture alluded to above (Hillas et al.
1998b).  A proof of detection of the IR background radiation through a
TeV photon absorption requires a detailed study of the spectrum of TeV
blazars and their spectral variability.  However, the IR limits 
(Biller et al. 1998) that allow for uncertainties in spectral shape 
are unchanged by this work. 

In summary, we have shown that the TeV spectra of Markarian 421 and
Markarian 501 differ significantly, the latter showing more curvature
and a harder spectral slope below 2 TeV.
Since the redshifts are almost identical, this difference can only be
attributed to physics intrinsic to the objects themselves, and it is
not inconsistent with a synchrotron-inverse Compton picture. 

\acknowledgments

We acknowledge the technical assistance of K. Harris and E. Roache. 
This research is supported by grants from the U.S. Department of Energy 
and by NASA, by PPARC in the UK and by Forbairt in Ireland.

\vfill\eject

\clearpage
\begin{deluxetable}{ccccccc}
\footnotesize
\tablecaption{Relative atmospheric transmission at SZA and LZA,
\label{tbl-1}}
\tablewidth{0pt}
\tablehead{
\colhead{} &
\colhead{} & \colhead{} &
\colhead{}  & \colhead{ Transmission:} \\
\colhead{Rayleigh} & \colhead{Aerosol} &
\colhead{$\rm O_{2} $} & \colhead{Ozone} & \colhead{ $\rm (z=0^{\circ}$)} &\colhead{$\rm (z=60^{\circ}$)} }
\startdata
  1        & 1  &  1  & 1  & 1.0 &  0.78  \nl
  1.03      & 1  &  1  & 1  & 0.99 &  0.77  \nl
  1        & 4  &  1  & 1  & 0.96 &  0.73  \nl 
  1        & 1  &  4  & 1  & 0.98 &  0.76  \nl
  1        & 1  &  1  & 4  & 0.93 &  0.74  \nl
\enddata
\end{deluxetable}

\clearpage

\figcaption[m42d]
{The collection areas for two different zenith angle ranges: $\rm
55^{\circ}-60^{\circ}$ (stars), and $\rm 20^{\circ}$ (bars)
for the 151 pixel camera of the Whipple telescope in 1997. It should
be noted that the 151 camera has increased the collection area at
$\rm 20^{\circ} $ relative to the older 109 pixel camera because of
its larger field of view.}

\figcaption[m42d]
{The energy spectrum of the Crab Nebula derived from large
zenith angle observations (open circles) in comparison with the energy
spectrum derived from small zenith angle (stars) data (Hillas et al. 1998a)
is shown above. The spectral fits to the two data-sets with a power-law
are consistent and are indicated by the solid and dashed lines. }

\figcaption[m42d]
{The energy spectra of Markarian~421 from the data of the
big flare on May 7 (stars) and a short flare 
on May 15 (rectangles), and LZA data from flaring states in
1995 (open circles) is shown above.  The flux of the LZA data
has been presented as 0.2 of the absolute flux.}

\figcaption[m42d]
{ The energy spectrum of Markarian~421 combining the data from the big
flare on May 7 (stars), the short flare on May 15 (rectangles) and a
flaring state in 1995 (open circles) at LZA is shown above.  The
absolute fluxes of the LZA data and the SZA data from May 15 have been
normalized to the big flare data.  }

\figcaption[m42d]
{The energy spectrum of Markarian~501 (LZA data only) in comparison to
the spectrum of the Crab Nebula derived with the same method is shown
above. A power-law fit to the Markarian 501 data gives a $\rm \chi^{2}
= 14.7$ for 5 degrees of freedom (probability of 0.015).}

\figcaption[m42d]
{The energy spectrum of Markarian~501 using 15 hours of SZA data
(stars) and 5 hours of LZA (open circles) data is shown above. For this plot the 
absolute flux of the LZA data has been normalized to the SZA data.  
Both a simple power-law (solid line) and a curved fit (dashed line) 
are shown in the figure. }
 
\figcaption[m42d]
{The energy spectrum of Markarian~501 (filled circles) and Markarian 421
(open stars) are compared above.  The solid line is a curved
fit to the Markarian 421 fluxes and the dotted line is a curved fit to
the Markarian 501 fluxes. } 

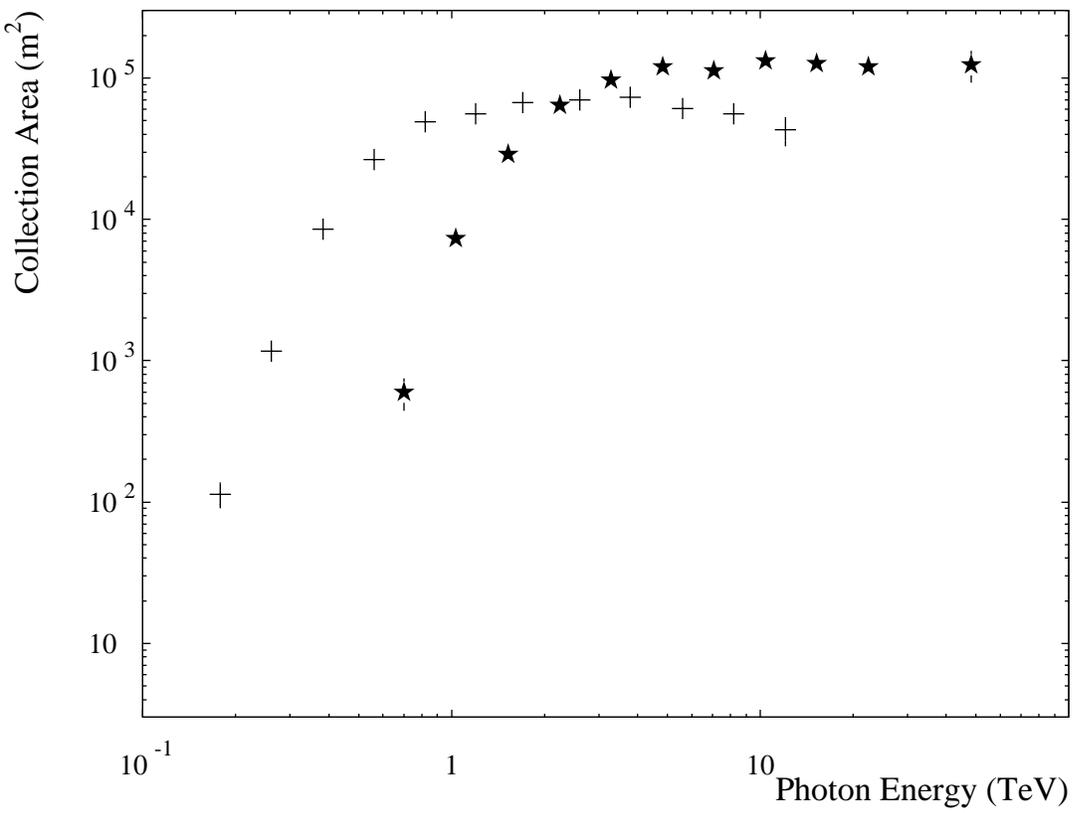
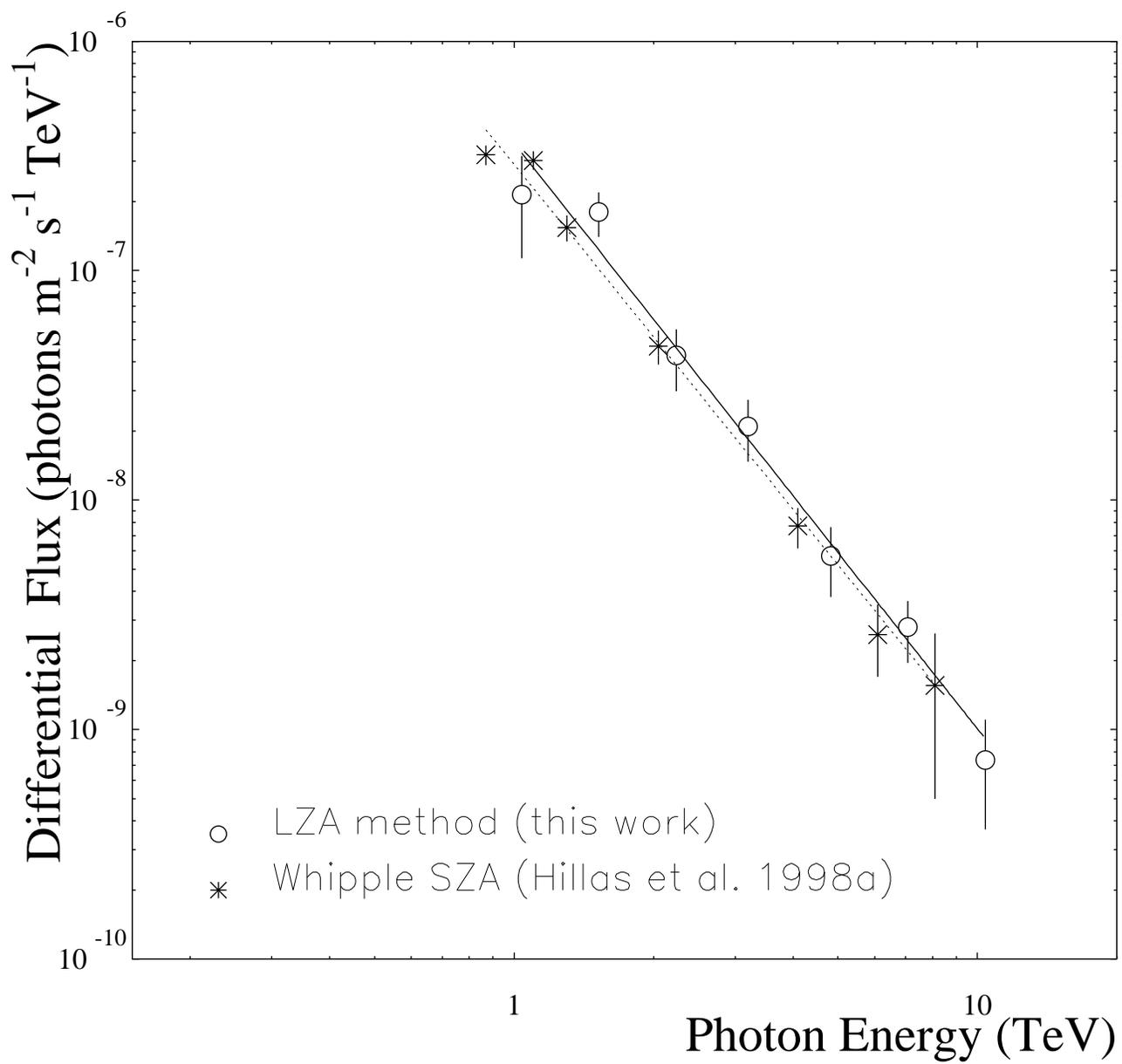
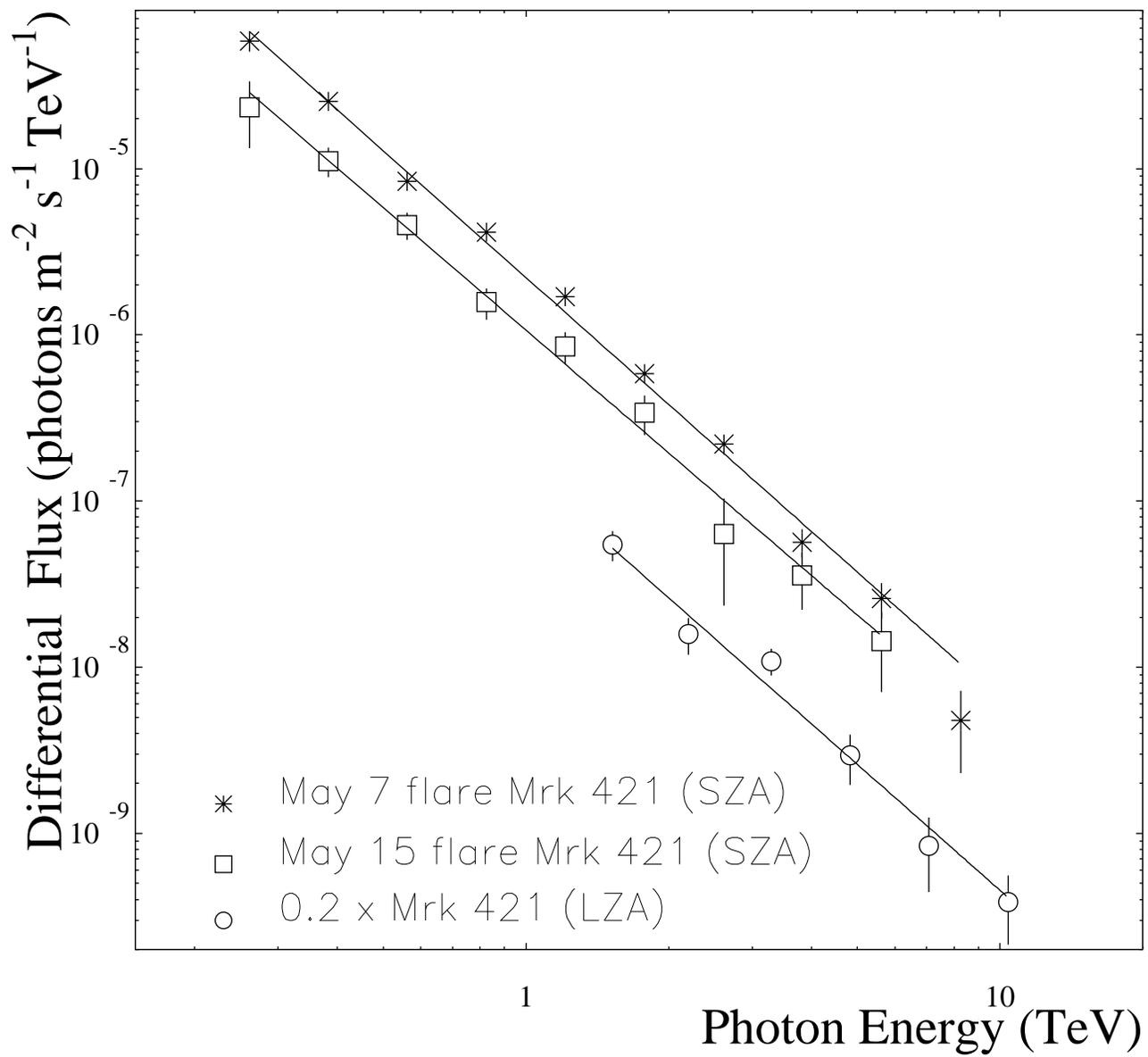
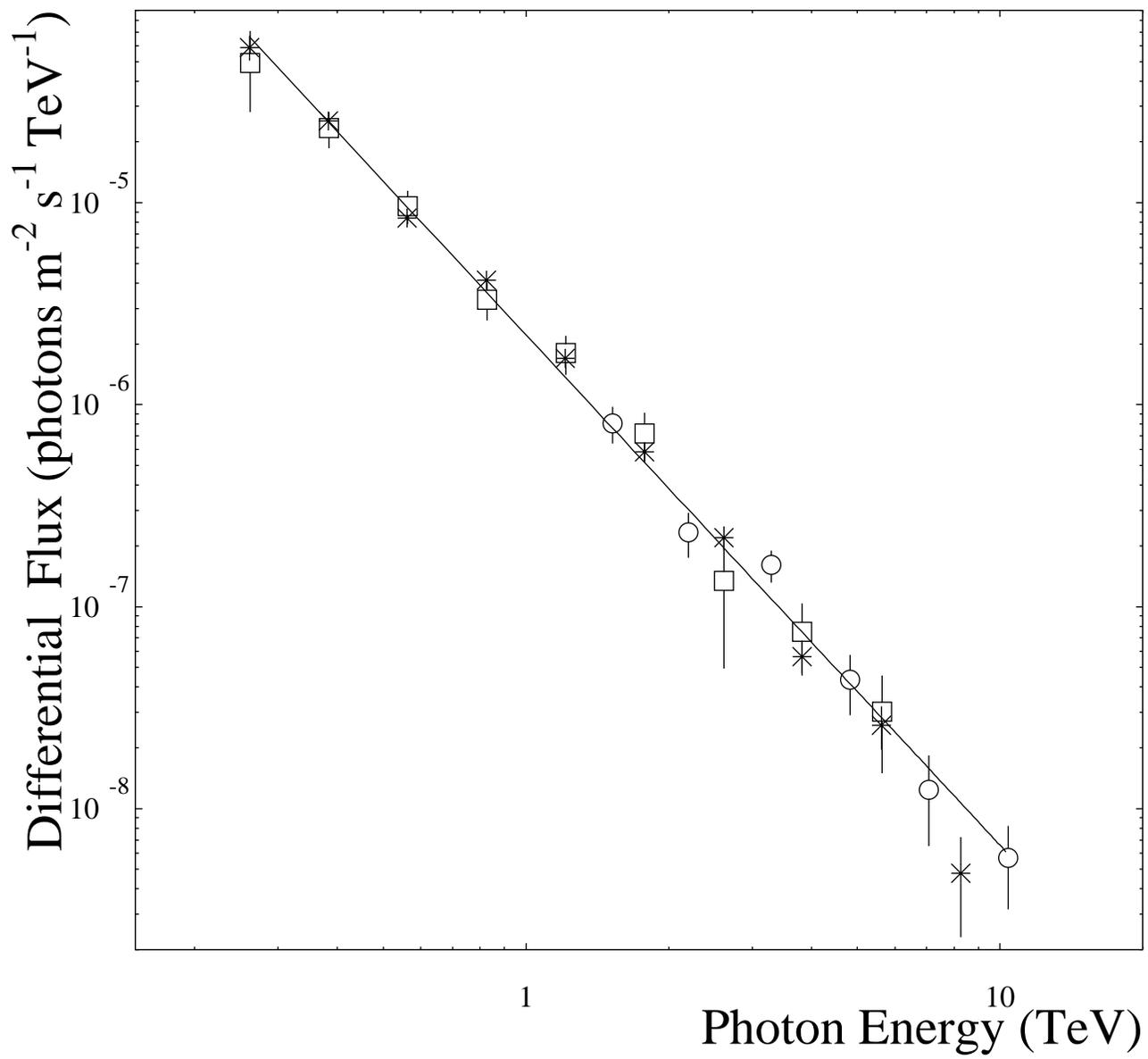
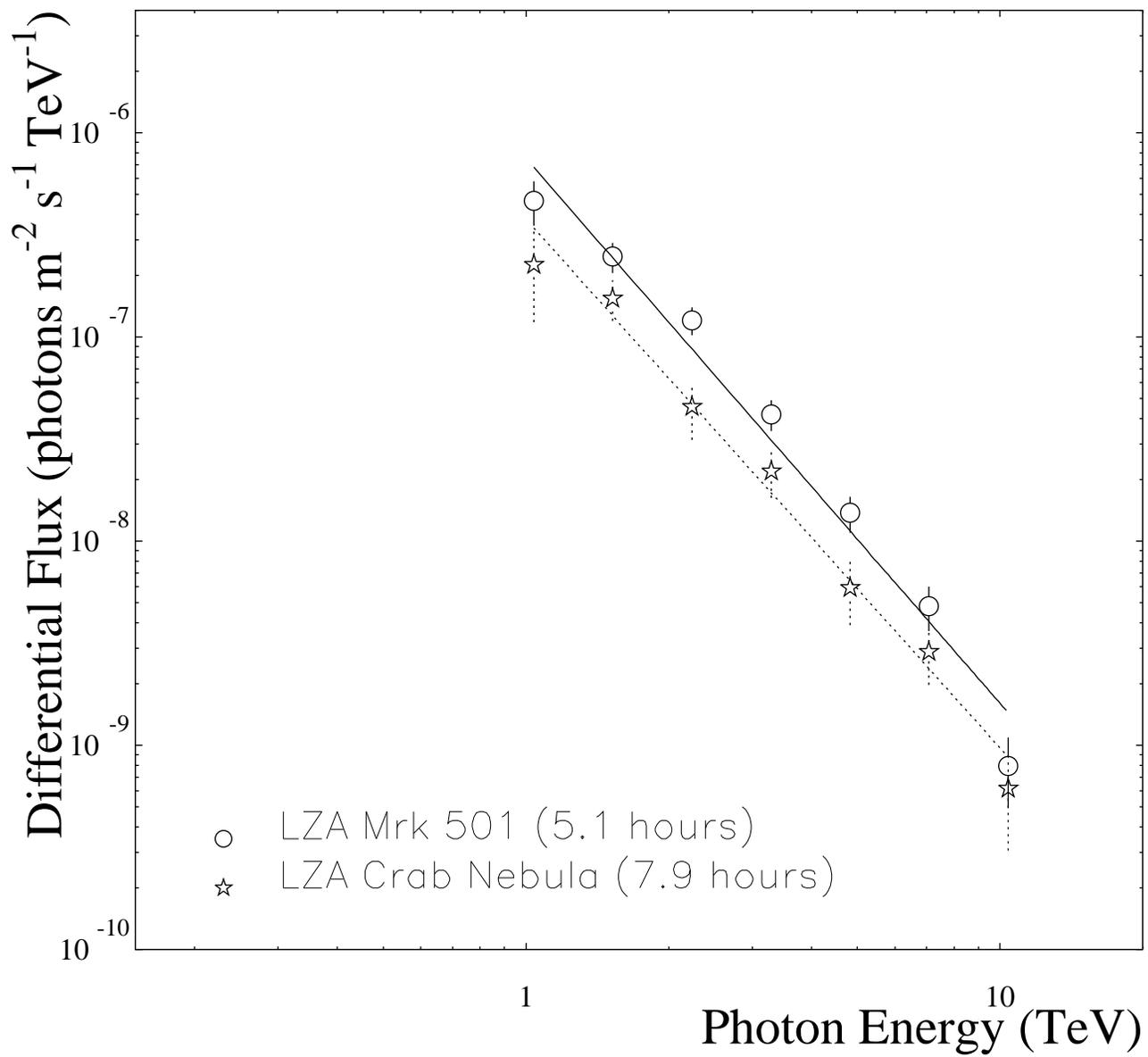
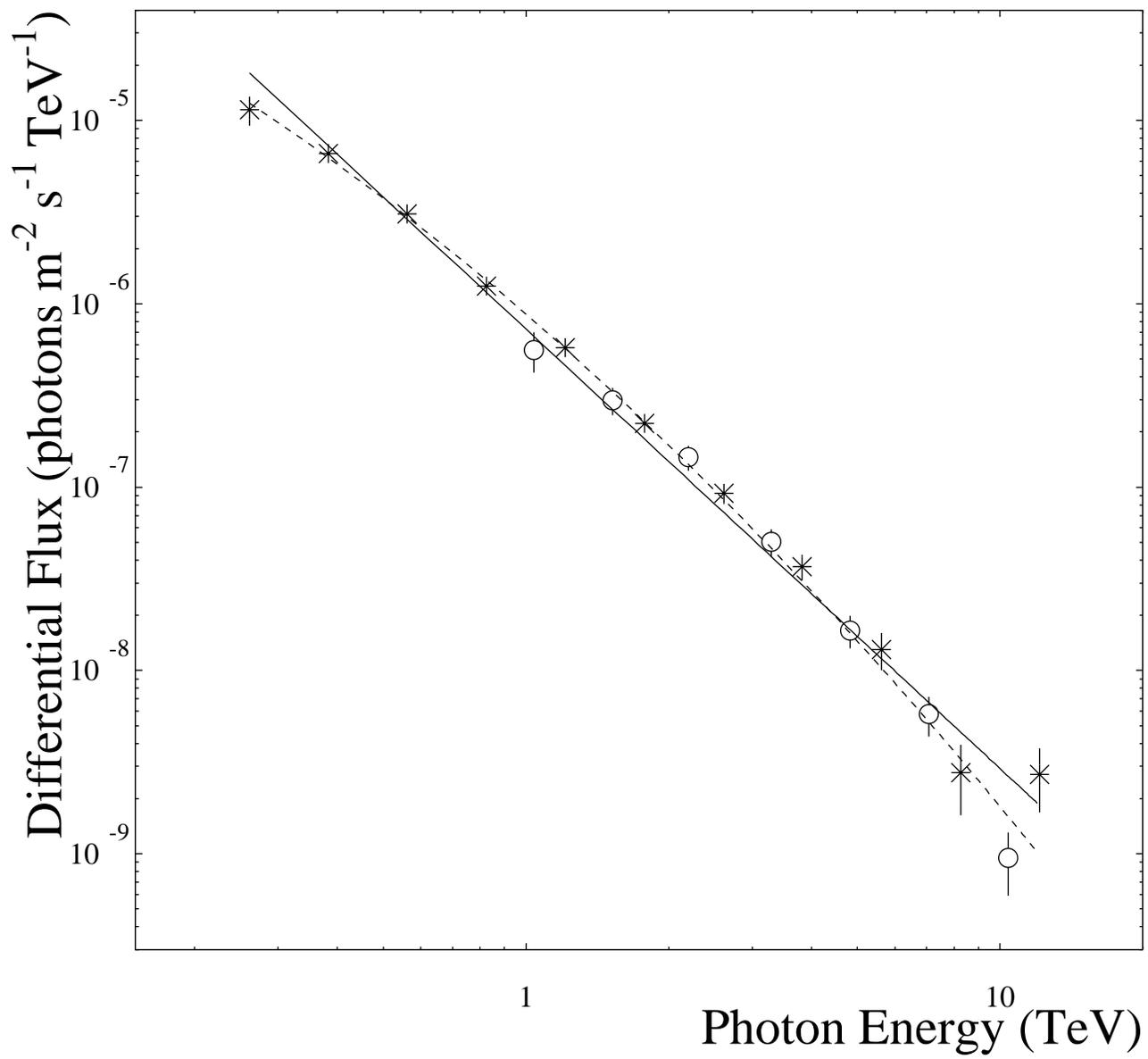
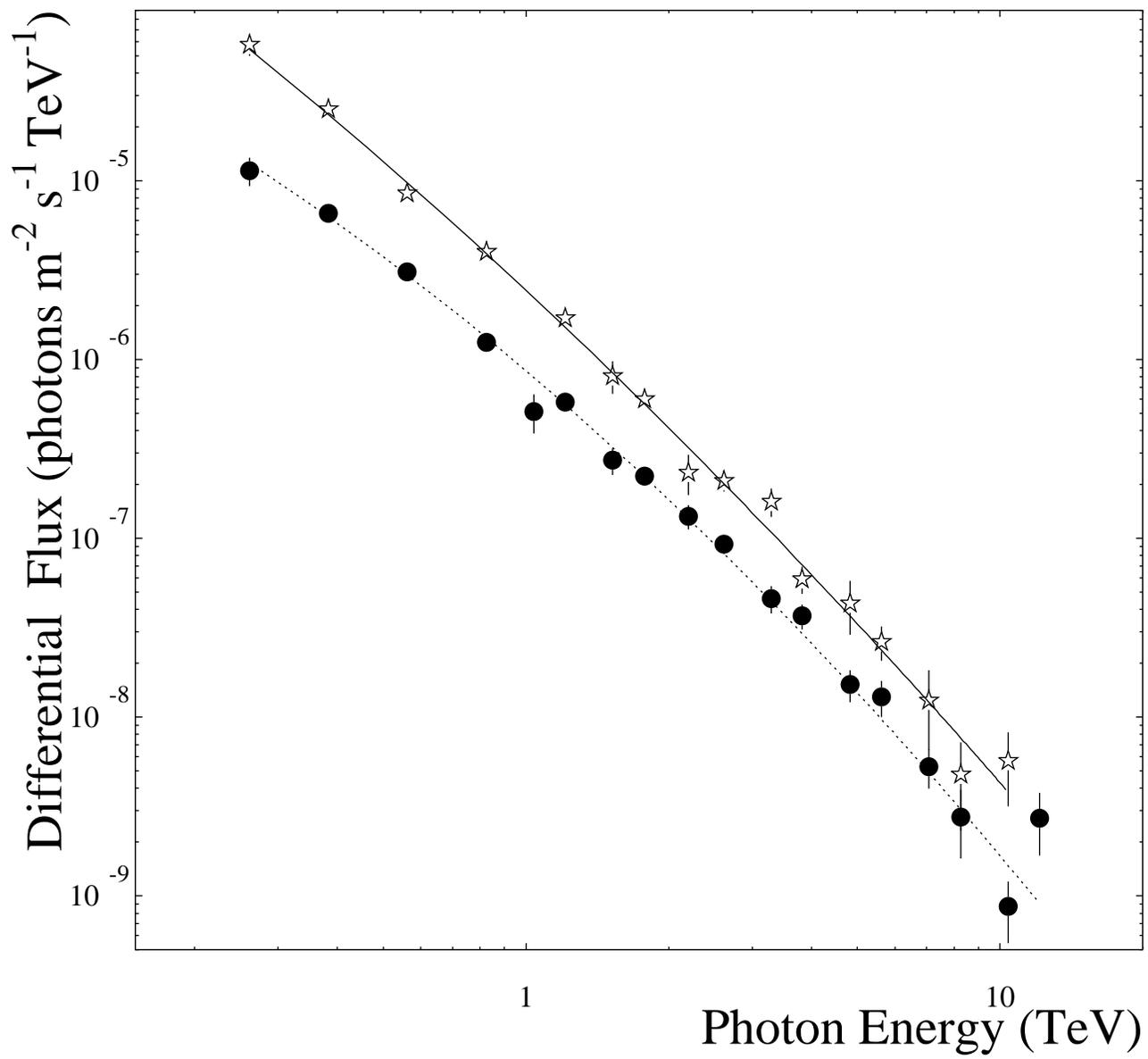
\end{document}